\documentstyle[11pt,newpasp,twoside,epsf]{article}
\def\emphasize#1{{\sl#1\/}}

\def\edcomment#1{\iffalse\marginpar{\raggedright\sl#1\/}\else\relax\fi}
\reversemarginpar

\begin{document}
\title{Radial-Velocity Monitoring of Members and Candidate Members of the TW Hydrae Association}
\author{Guillermo Torres}
\affil{Harvard-Smithsonian Center for Astrophysics, 60 Garden St., Mail Stop 20, Cambridge, MA 02138, USA}
\author{Ralph Neuh\"auser}
\affil{Max-Planck-Institut f\"ur extraterrestrische Physik, Giessenbachstrasse 1, 85740 Garching, Germany}
\author{David W. Latham}
\affil{Harvard-Smithsonian Center for Astrophysics, 60 Garden St., Mail Stop 20, Cambridge, MA 02138, USA}

\begin{abstract} We present our spectroscopic measurements of the
radial velocity, effective temperature, and projected rotational
velocity of several of the known members of the TW Hya association, as
well as measurements for candidate members selected on the basis of
their X-ray or kinematic properties.  A number of our targets turn out
to be binaries, but most are non-members. The radial velocities for
some of the other candidates support the conclusion that they are
kinematically associated with the group, although further observations
are required to show that they are indeed pre-main sequence objects.
\end{abstract}

\section{Introduction}

The TW~Hya association, a loose grouping of very young stars that does
not appear to be related to any nearby molecular cloud material, has
been the subject of considerable attention in recent years. The
prototype of the association, TW Hya itself, was the first recognized
example of an isolated T Tauri star. The first few members of the
association were discovered more or less by accident, and were only
later recognized to be related. Aside from these, the systematic
search for additional stars belonging to the group has relied on their
X-ray properties (such as their detection as a ROSAT source), or their
similar kinematics (proper motions). 

Follow-up studies have been made in many cases to establish the youth
of the candidate members, for example by the presence of strong
Lithium $\lambda$6708 absorption, H$\alpha$ emission, infrared excess,
etc. To date there are 19 objects recognized as true members in the
published literature (see, e.g., Kastner et al.\ 1997; Webb et al.\
1999; Sterzik et al.\ 1999; Zuckerman et al.\ 2001), along with many
other potential members that need to be studied further. One of the
goals of the present work is to investigate the kinematics of the
candidate members, but in the direction \emphasize{orthogonal} to
previous studies that have focussed only on proper motions. We present
measurements along the line of sight (radial velocities), which are
complementary to the proper motions and help define the space motion
of these objects. Another goal of this project is to detect any
spectroscopic binaries that might be present through repeated
measurements of the radial velocities of our targets, and to determine
their orbits. 

\section{Sample and observations}

Our observations were obtained using nearly identical echelle
spectrographs on the MMT prior to conversion (Mt.\ Hopkins, AZ), the
1.5m Tillinghast reflector at the F.\ L.\ Whipple Observatory (also on
Mt.\ Hopkins, AZ), and on the 1.5m Wyeth reflector at the Oak Ridge
Observatory (Harvard, MA). A single echelle order was recorded with a
photon-counting Reticon diode array, at a resolving power of
$\lambda/\Delta\lambda = 35,\!000$ and a central wavelength of
5187\AA. The spectral coverage is 45\AA. We report here on the results
based on some 300 spectra, and an additional 150 archival spectra from
the CfA echelle database. 

Table 1 presents the sample of stars studied in this program,
including for completeness HD~98800 (TWA-4), which was observed
several years ago with the same instrumental setup and was discovered
to be a quadruple system (Torres et al.\ 1995). The second column in
the table gives the ``official" designation within the association,
and the last column collects the available measurements of the Li
$\lambda$6708 equivalent width for these stars. 

\vskip -5pt		
\scriptsize

\begin{table}
\caption{Summary of Optical Properties.}
\vspace{5pt}
\begin{tabular}{rllcccc@{~~}c}
\tableline
\noalign{\vspace{3pt}}
   &          &                          &    R.A.      &     Dec.      &   $V$          &                & Li $\lambda$6708 eq. \\
   & TWA \#   &  Other name              &  (J2000)     &   (J2000)     &  (mag)         &  SpT           & width (\AA)    \\ 
\noalign{\vspace{3pt}}
\tableline
\noalign{\vspace{3pt}}
 1 &          &  HIP 48273               &  09:50:30.1  &  $+$04:20:37  &  \phantom{2}6.24      &  F6            &    ...             \\
 2 &          &  TYC 6604-0118-1         &  09:59:08.4  &  $-$22:39:35  &  10.09         &  K2            &  0.143             \\
 3 &          &  HIP 50796               &  10:22:18.0  &  $-$10:32:15  &  10.80         &    ...         &    ...             \\
 4 &          &  HIP 53486               &  10:56:31.0  &  $+$07:23:19  &  \phantom{2}7.37      &  K0            &    ...             \\
 5 &          &  RXJ1100.0$-$3813        &  11:00:02.4  &  $-$38:13:20  &  12.29         &    ...         &    ...             \\
 6 &          &  HD~95490                &  11:00:51.2  &  $-$35:33:38  &  \phantom{2}8.79      &  F7      &  0.15\phantom{2}              \\
 7 &  TWA-1   &  TW Hya                  &  11:01:52.0  &  $-$34:42:16  &  10.92         &  K8e   &  0.43\phantom{2} \\
 8 &  TWA-2A  &  CD$-29\deg$8887A     &  11:09:13.9  &  $-$30:01:39  &  11.07         &  M2e &  0.52\phantom{2}        \\
 9 &          &  RXJ1109.7$-$3907        &  11:09:40.1  &  $-$39:06:48  &  10.58         &  G3            &  0.190             \\
10 &  TWA-3A  &  Hen 3-600A              &  11:10:28.0  &  $-$37:31:53  &  12.04         &  M4e       &  0.57\phantom{2}       \\
11 &          &  HD~97131                &  11:10:34.2  &  $-$30:27:19  &  \phantom{2}9.01      &  F2        &  0.03\phantom{2}              \\
12 &          &  CD$-37\deg$7097      &  11:12:42.7  &  $-$38:31:04  &  10.24         &  F5            &    ...             \\
13 &          &  RXJ1115.1$-$3233        &  11:15:06.9  &  $-$32:32:46  &  12.36         &    ...         &    ...             \\
14 &  TWA-12  &  RXJ1121.1$-$3845        &  11:21:05.5  &  $-$38:45:17  &  12.85         &  M2            &  0.530             \\
15 &  TWA-13A &  RXJ1121.3$-$3447N       &  11:21:17.3  &  $-$34:46:47  &  11.46         &  M2e       &  0.650             \\
16 &  TWA-13B &  RXJ1121.3$-$3447S       &  11:21:17.5  &  $-$34:46:51  &  12.00         &  M1e       &  0.51\phantom{2} \\
17 &  TWA-4A  &  HD 98800A               &  11:22:05.3  &  $-$24:46:40  &  \phantom{2}9.41      &  K5            &  0.425             \\
18 &  TWA-4B  &  HD 98800B               &  11:22:05.3  &  $-$24:46:39  &  \phantom{2}9.94      &    ...         &  0.335$+$0.450   \\
19 &  TWA-5A  &  CD$-33\deg$7795A     &  11:31:55.4  &  $-$34:36:27  &  11.37         &  M3      &  0.55\phantom{2}        \\
20 &  TWA-9A  &  CD$-36\deg$7429A     &  11:48:24.3  &  $-$37:28:49  &  11.26         &  K5            &  0.46\phantom{2}              \\ 
\noalign{\vspace{2pt}}
\tableline\tableline
\end{tabular}
\end{table}
\normalsize

All recognized members are seen to have strong Li in absorption, while
some of the other objects (proposed by a number of authors as
candidate members) have much weaker Li. Not all of the potential
members have estimates of this crucial youth indicator, so their
status is unclear. 

\section{Results}

Radial velocities were obtained using standard cross-correlation
techniques (the XCSAO task running under IRAF), as described, e.g., by
Latham (1992). The templates were taken from an extensive library of
synthetic spectra based on the latest model atmospheres by Kurucz
(Morse \& Kurucz, in preparation). In addition, we derived effective
temperatures ($T_{\rm eff}$) and projected rotational velocities ($v
\sin i$) for all our stars by comparison with the synthetic spectra. A
number of objects have turned out to be double-lined. For these we
used the two-dimensional cross-correlation algorithm known as TODCOR
(Zucker \& Mazeh 1994), and in most cases we were able to derive the
$T_{\rm eff}$ and $v \sin i$ for both components. 

Table 2 lists our results for the temperatures (K) and rotational
velocities (km s$^{-1}$). A comparison with $T_{\rm eff}$ values
derived from the spectral types (adopting the calibration by de Jager
\& Nieuwenhuijzen 1987) and other sources shows generally good
agreement. The agreement for the cooler stars is poorer due to
limitations in the model atmospheres used to compute the synthetic
spectra, which do not include several key molecular opacity sources.
Our measures of $v \sin i$ are also fairly consistent with other
determinations. 

\vskip -5pt	
\footnotesize

\begin{table}
\caption{Effective temperatures and rotational velocities.}
\vspace{5pt}
\begin{tabular}{rll@{~~~}c@{~~}c@{~}c@{}c@{}c@{}c}
\tableline
\noalign{\vspace{3pt}}
 &  &  &  & T$_{\rm eff}$ & T$_{\rm eff}$ & T$_{\rm eff}$ & $v \sin i$
 & $v \sin i$ \\
 & TWA \# & Other names & SpT & (CfA) & (SpT) & (Other) & (CfA) & (Other) \\
\noalign{\vspace{3pt}}
\tableline
\noalign{\vspace{3pt}}
 1 &          &  HIP 48273           &  F6            &  6300/6150         &  6530  &         & 22/22            &        \\
 2 &          &  TYC 6604-0118-1     &  K2            &  5050/4850         &  4840  &         & 19/15            & 19     \\
 3 &          &  HIP 50796           &    ...         &  4750              &  ...   &         & 8                &        \\
 4 &          &  HIP 53486           &  K0            &  5050              &  5150  &         & \phantom{:}1:    &        \\
 5 &          &  RXJ1100.0$-$3813    &    ...         &  5750              &  ...   &         & 35               &        \\
 6 &          &  HD~95490            &  F7      &  6500              &  6430  &         & 13               &        \\
 7 &  TWA-1   &  TW Hya              &  K8e   &  4150              &  4150  &  4150   & \phantom{2}4            & 10,~15,~14 \\
 8 &  TWA-2A  &  CD$-29\deg$8887A    &  M2e &  \phantom{:}4050:  &  3690  &         & 13               & 15     \\
 9 &          &  RXJ1109.7$-$3907    &  G3            &  5900              &  5710  &  5800   & 27               & 23     \\
10 &  TWA-3A  &  Hen 3-600A          &  M4e       &  \phantom{:}4750:  &  3350  &         & \phantom{:}20:   & 15     \\
11 &          &  HD~97131            &  F2        &  6750/6750         &  7180  &         & 12/16            &        \\
12 &          &  CD$-37\deg$7097     &  F5            &  6250              &  6650  &         & 14               &        \\
13 &          &  RXJ1115.1$-$3233    &    ...         &  \phantom{:}5750/5250: &  ...   &         & \phantom{:}25/0:\phantom{2} &        \\
14 &  TWA-12  &  RXJ1121.1$-$3845    &  M2            &  \phantom{:}4000:      &  3520  &  3600   & 15               & 21     \\
15 &  TWA-13A &  RXJ1121.3$-$3447N   &  M2e       &  \phantom{:}4150:      &  3520  &  3600   & 12               & 16,~10 \\
16 &  TWA-13B &  RXJ1121.3$-$3447S   &  M1e        &  \phantom{:}4100:      &  3660  &  3800   & 12               & 16,~12 \\
17 &  TWA-4A  &  HD 98800A           &  K5            &                    &  4410  &  4350   &                  & 5      \\
18 &  TWA-4B  &  HD 98800B           &    ...         &                    &  ...   &  4250/3700   &             & 3/2    \\
19 &  TWA-5A  &  CD$-33\deg$7795A    &  M3      &  \phantom{:}4050:  &  3490  &         & \phantom{:}36:   & 58     \\
20 &  TWA-9A  &  CD$-36\deg$7429A    &  K5            &  4350              &  4410  &         & 11               &        \\
\noalign{\vspace{2pt}}
\tableline\tableline
\end{tabular}
\end{table}
\normalsize

In addition to the well known quadruple system HD~98800, multiple
measurements of the radial velocity of our other targets have revealed
several binaries (some double-lined) with short orbital periods
(RXJ1115.1$-$3233, TYC~6604-0118-1, HIP~48273, RXJ1100.0$-$3813) and
one double-lined triple system (HD 97131). None of these appear to be
true members of the TW Hya association, based on their Li abundance or
kinematics (radial velocity of the center of mass).  Details on these
orbital solutions will be reported in a forthcoming paper. 

In Table 3 we list the mean heliocentric radial velocities of all our targets. For
the multiple systems we give the center-of-mass velocity. Our
measurements are consistent with other determinations from the
literature, but much more precise. Except for the case of Hen~3-600A
(TWA-3A) and CD$-$33\deg7795A (TWA-5A), which are double-lined or
possibly triple-lined but have not yet had their orbits determined
(see also Webb et al.\ 1999), the other recognized members of the
association have similar radial velocities.  In the case of HD~98800
we have assumed that the average velocity of components A and B
corresponds to the center of mass of the quadruple system. Similarly
for RXJ1121.3$-$3447 (TWA-13), which is a visual binary. The mean
velocity of TWA-1, TWA-2, TWA-4, TWA-9, TWA-12, and TWA-13 is $+11.30
\pm 0.56$ km~s$^{-1}$. 

\vskip -10pt
\scriptsize

\begin{table}
\caption{Radial velocity results (km s$^{-1}$).}
\vspace{5pt}
\begin{tabular}{rllcc@{}c@{}l}
\tableline
\noalign{\vspace{3pt}}
 &  &  &  & Mean $RV$ & Mean $RV$ &  \\
 & TWA \# & Other names & N$_{\rm obs}$ & (CfA) & (Other) & Remarks \\
\noalign{\vspace{3pt}}
\tableline
\noalign{\vspace{3pt}}
 1 &          &  HIP 48273              & 112       &  $+$16.249~$\pm$~0.071\phantom{$+1$}     &                             & Binary (orbit) \\
 2 &          &  TYC 6604-0118-1        & \phantom{1}34    &  $+$26.96~$\pm$~0.24\phantom{$+1$}       &                             & Binary (orbit) \\
 3 &          &  HIP 50796              & \phantom{12}2 &  $+$13.1~$\pm$~1.0\phantom{$+1$}         &                             & \\
 4 &          &  HIP 53486              & \phantom{12}6 &  $+$5.54~$\pm$~0.29\phantom{$+$}         &                             & \\
 5 &          &  RXJ1100.0$-$3813       & \phantom{1}18    &  $+$2.03~$\pm$~0.92\phantom{$+$}         &                             & Binary (orbit) \\
 6 &          &  HD~95490               & \phantom{12}4 &  $-$7.51~$\pm$~0.17\phantom{$-$}         &                             & \\
 7 &  TWA-1   &  TW Hya                 & \phantom{12}5 &  $+$12.92~$\pm$~0.23\phantom{$+1$}       & $+$13.5~$\pm$~1.5\phantom{$+1$} & \\
 8 &  TWA-2A  &  CD$-$29$\deg$8887A     & \phantom{12}6 &  $+$11.20~$\pm$~0.32\phantom{$+1$}       & $+12$\phantom{$+$}              & \\
 9 &          &  RXJ1109.7$-$3907       & \phantom{12}7 &  \phantom{:}$-$1.2~$\pm$~1.1:\phantom{$+$}   & $-$2.0~$\pm$~1.1\phantom{$-$}   & VAR? \\
10 &  TWA-3A  &  Hen 3-600A             & \phantom{1}14    &  \phantom{:}$+$6.9~$\pm$~1.7:\phantom{$+$}   & $+14$\phantom{$+$}              & Double-lined? \\
11 &          &  HD~97131               & \phantom{1}37    &  $-$27.10~$\pm$~0.26\phantom{$-2$}       &                             & Triple (orbit) \\
12 &          &  CD$-37\deg$7097        & \phantom{12}3 &  $-$8.19~$\pm$~0.26\phantom{$-$}         &                             & \\
13 &          &  RXJ1115.1$-$3233       & \phantom{1}11    &  $-$10.0~$\pm$~1.1\phantom{$-1$}         &                             & Binary (orbit)\\
14 &  TWA-12  &  RXJ1121.1$-$3845       & \phantom{12}2 &  $+$12.23~$\pm$~0.60\phantom{$+1$}       & $+$10.9~$\pm$~1.1\phantom{$+1$} & \\
15 &  TWA-13A &  RXJ1121.3$-$3447N      & \phantom{12}4 &  $+$11.72~$\pm$~0.61\phantom{$+1$}       & $+$10.5~$\pm$~1.2\phantom{$+1$} & \\
16 &  TWA-13B &  RXJ1121.3$-$3447S      & \phantom{12}4 &  $+$12.41~$\pm$~0.48\phantom{$+1$}       & $+$12.0~$\pm$~1.2\phantom{$+1$} & \\
17 &  TWA-4A  &  HD 98800A              & 152       &  $+$12.75~$\pm$~0.10\phantom{$+1$}       &                             & Binary (orbit) \\
18 &  TWA-4B  &  HD 98800B              & 152       &  $+$5.73~$\pm$~0.14\phantom{$+$}         &                             & Binary (orbit) \\
19 &  TWA-5A  &  CD$-$33$\deg$7795A     & \phantom{1}26    &  \phantom{:}$+$6.9~$\pm$~2.0:\phantom{$+$}   & $+14$\phantom{$+$}              & Double-lined? \\
20 &  TWA-9A  &  CD$-$36$\deg$7429A     & \phantom{1}10    &  $+$10.17~$\pm$~0.36\phantom{$+1$}       &                             &\\
\noalign{\vspace{2pt}}
\tableline\tableline
\end{tabular}
\end{table}
\normalsize

Though it is tempting to assign this mean radial velocity to the group
as a whole, and to then use it as a criterion to accept or reject other
candidate members, reality is more complex and various kinematical
studies have shown that there is probably a gradient in the radial
velocity across the large sky area covered by this association
(several tens of degrees). One of such studies, by Makarov \&
Fabricius (2001), has modeled the kinematics of the association as a
moving group using a variant of the convergent-point method.  The
authors used proper motions from the Tycho-2 catalogue along with
Hipparcos parallaxes for the few members which have them, and searched
an area of more than 3000~deg$^2$ for additional members with the same
motion that are also X-ray sources from the ROSAT Bright Source
Catalog. For each star they computed a ``kinematical" distance as well
as the predicted radial velocity. Comparing the latter with the few
measurements available to the authors from the literature for the
classical members, they found it necessary to include an expansion
term in their model, which not only is not unexpected for this group,
but also gives a dynamical age of 8.3~Myr, in good agreement with
other estimates from pre-main sequence evolutionary tracks ($\sim
10$~Myr). The internal velocity dispersion they derived is 0.8
km~s$^{-1}$, and the depth of the group is significant (tens of
parsecs). 

Makarov \& Fabricius (2001) produced a list of 23 additional
candidates, of which we have observed several (see Table 4).  Our
radial velocity measurements allow us to test these objects for
membership. The known members of the association (designation in
column 2) have radial velocities ($RV_{\rm obs}$) very close to the
predicted values ($RV_{\rm pred}$) in most cases, but two of the other
stars (HIP 48273 and TYC 6604-0118-1), which happen to be double-lined
binaries, do not. On the other hand, the velocity for HIP 50796 agrees
perfectly with the prediction. HIP 53486 is an especially interesting
case because it the nearest candidate member (17 pc), and its
predicted distance ($D_{\rm kin}$) agrees with the value measured by
Hipparcos ($D_{\rm trig}$).  It has the lowest expected radial
velocity ($+3.7$ km~s$^{-1}$), and we do indeed measure a low value
($+5.5$ km~s$^{-1}$).  Further evidence that some of these new
candidates may be true members is given by the fact that, unbeknownst
to Makarov \& Fabricius, one of their stars (HIP 57524, too faint for
us and not shown in Table 4) is actually TWA-19, which is Li-rich.
Follow-up observations of all these candidate are necessary to
establish their youth, and our colleagues at this Workshop are
encouraged to do so.

\scriptsize

\begin{table}
\caption{Candidate members from the kinematical study by Makarov \& Fabricius (2001) with measured RVs from our work.}
\vspace{5pt}
\begin{tabular}{l@{ }c@{ }c@{~~}c@{~~}c@{~~~}c@{~~}cc@{~~}c}
\tableline
\noalign{\vspace{3pt}}
     &        & R.A. & Dec. & $\mu$  & $D_{\rm trig}$ & $D_{\rm kin}$ & $RV_{\rm pred}$ & $RV_{\rm obs}$ \\
Name & TWA \# & (2000) & (2000) & (mas/yr) & (pc) & (pc) & (km~s$^{-1}$) & (km~s$^{-1}$) \\
\noalign{\vspace{3pt}}
\tableline
\noalign{\vspace{3pt}}
HIP 53911       & 1 & 11:01:51.9 & $-$34:42:17 & \phantom{2}75.4 &  56.4 & 57.1 & $+$12.7           & $+$12.9\phantom{:}   \\
TYC 7201-0027-1 & 2 & 11:09:13.8 & $-$30:01:40 & \phantom{2}92.6 &       & 47.1 & $+$10.6           & $+$11.2\phantom{:}   \\
HIP 55505       & 4 & 11:22:05.3 & $-$24:46:40 & \phantom{2}96.8 &  46.7 & 45.7 & \phantom{1}$+$9.1 & \phantom{1}$+$9.2\phantom{:}   \\
TYC 7223-0275-1 & 5 & 11:31:55.3 & $-$34:36:28 & \phantom{2}86.7 &       & 50.9 & $+$10.0           & \phantom{1}$+$6.9:  \\
HIP 57589       & 9 & 11:48:24.2 & $-$37:28:49 & 158.1           &  50.3 & 76.3 & $+$12.6           & $+$10.2\phantom{:}   \\
HIP 48273       &   & 09:50:30.1 & $+$04:20:37 & 157.9           &  45.9 & 26.6 & $+$10.7           & $+$16.2\phantom{:}   \\
TYC 6604-0118-1 &   & 09:59:08.4 & $-$22:39:35 & 163.7           &       & 62.8 & $+$16.5           & $+$27.0\phantom{:}   \\
HIP 50796       &   & 10:22:18.0 & $-$10:32:16 & 179.1           &  34.0 & 53.8 & $+$13.1           & $+$13.1\phantom{:}   \\
HIP 53486       &   & 10:56:30.8 & $+$07:23:18 & 268.1           &  17.6 & 16.7 & \phantom{1}$+$3.7 & \phantom{1}$+$5.5\phantom{:}   \\
\noalign{\vspace{2pt}}
\tableline\tableline
\end{tabular}
\end{table}


\normalsize

\begin{references}

\reference de Jager, C. \& Nieuwenhuijzen, H. 1987, \aap, 177, 217

\reference Kastner, J.\ H., Zuckerman, B., Weintraub, D.\ A., \&
Forveille, T. 1997, Science, 277, 67

\reference Latham, D.\ W. 1992, in IAU Coll.\ 135, Complementary
Approaches to Double and Multiple Star Research, eds.\ H.\ A.\
McAlister \& W.\ I.\ Hartkopf (San Francisco: ASP), 110

\reference Makarov, V.\ V., \& Fabricius, C. 2001, \aap, 368, 866

\reference Sterzik, M.\ F., Alcal\'a, J.\ M., Covino, E., \& Petr, M.\
G. 1999, \aap, 346, L41

\reference Torres, G., Stefanik, R.\ P., Latham, D.\ W., \& Mazeh, T.
1995, \apj, 452, 870

\reference Webb, R.\ A., Zuckerman, B., Platais, I., Patience, J.,
White, R.\ J., Schwartz, M.\ J., \& McCarthy, C. 1999, \apj, 512, L63

\reference Zucker, S., \& Mazeh, T. 1994, \apj, 420, 806

\reference Zuckerman, B., Webb, R.\ A., Schwartz, M., \& Becklin, E.\
E. 2001, \apj, 549, L233

\end{references}
\end{document}